\documentclass[12pt]{iopart}
\usepackage{iopams}
\usepackage{graphicx}
\usepackage{color}
\usepackage{amsfonts}
\input amssym.def 
\input amssym

\begin{document}

A following up on the FQXi Essay Contest - It from Bit or Bit from It? June 2013

\title[]{ Drawing quantum contextuality with \lq dessins d'enfants'}

\author{ Michel Planat}

\vspace*{.1cm}
\address{$^1$Institut FEMTO-ST, CNRS, 32 Avenue de
l'Observatoire, 25044 Besan\c con, France.}
\ead{michel.planat@femto-st.fr}

\begin{abstract}

In the standard formulation of quantum mechanics, there exists an inherent feedback of the measurement setting on the elementary object under scrutiny. Thus one cannot assume that an \lq element of reality' prexists to the measurement and, it is even more intriguing that unperformed/counterfactual observables enter the game. This is called quantum contextuality. Simple finite projective geometries are a good way to picture the commutation relations of quantum observables entering the context, at least for systems with two or three parties. In the essay, it is further discovered a mathematical mechanism for \lq drawing' the contexts. The so-called \lq dessins d'enfants' of the celebrated mathematician Alexandre Grothendieck feature group, graph, topological, geometric and algebraic properties of the quantum contexts that would otherwise have been \lq hidden' in the apparent randomness of measurement outcomes.

\end{abstract}

\section{Introduction}

The motivation for being interested in the topic of quantum contextuality dates back the celebrated Bohr-Einstein dialogue about the fundamental nature of quantum reality. The first sentence of the Einstein-Podolski-Rosen (EPR) paper \cite{Einstein1935} is as follows 

{\it If, without in any way disturbing a system, we can predict with certainty (i.e., with probability equal to unity) the value of a physical quantity, then there exists an element of physical reality corresponding to that physical quantity.}

\noindent
and the last sentence in Bohr's \cite{Bohr1935} reply is

{\it I should like to point out, however, that the named criterion contains an essential ambiguity when it is applied to problems of quantum mechanics. It is true that in the measurements under consideration any direct mechanical interaction of the system and the measuring agencies is excluded, but a closer examination reveals that the procedure of measurements has an essential influence on the conditions on which the very definition of the physical quantities in question rests. Since these conditions must be considered as an inherent element of any phenomenon to which the term \lq\lq physical reality" can be unambiguously applied, the conclusion of the above mentioned authors would not appear to be justified.}

\noindent
In a recent essay, taking into account the work of Bell about non-locality \cite{Bell1964} and further important papers by Gleason, Kochen-Specker \cite{Peres,Koch1967} 
and Mermin \cite{Mermin1993}, I arrived at the conclusion that further progress about the elusive {\it elements of reality}, or rather the {\it elements of knowledge}, can be performed by resorting to Grothendieck's {\it dessins d'enfants} as summarized in the note \cite{PlanatFQXi} and the paper \cite{Dessins2013}
 intended to illustrate Wheeler's {\it it from bit} perspective \cite{Wheeler}. In Grothendieck's words  \cite[(a), Vol. 1]{Schneps1}

\noindent
{\it The demands of university teaching, addressed to students (including
those said to be “advanced”) with a modest (and frequently less than modest)
 mathematical baggage, led me to a Draconian renewal of the themes
of reflection I proposed to my students, and gradually to myself as well.
It seemed important to me to start from an intuitive baggage common to
everyone, independent of any technical language used to express it, and
anterior to any such language – it turned out that the geometric and topological
 intuition of shapes, particularly two-dimensional shapes, formed such
a common ground. This consists of themes which can be grouped under the
general name of “topology of surfaces” or “geometry of surfaces”, it being
understood in this last expression that the main emphasis is on the topological
 properties of the surfaces, or the combinatorial aspects which form the
most down-to-earth technical expression of them, and not on the differential,
 conformal, Riemannian, holomorphic aspects, and (from there) on to
“complex algebraic curves”. Once this last step is taken, however, algebraic
geometry (my former love!) suddenly bursts forth once again, and this via
the objects which we can consider as the basic building blocks for all other
algebraic varieties. Whereas in my research before 1970, my attention was
systematically directed towards objects of maximal generality, in order to
uncover a general language adequate for the world of algebraic geometry,
and I never restricted myself to algebraic curves except when strictly necessary
 (notably in etale cohomology), preferring to develop “pass-key” techniques
 and statements valid in all dimensions and in every place (I mean,
over all base schemes, or even base ringed topoi...), here I was brought
back, via objects so simple that a child learns them while playing, to the
beginnings and origins of algebraic geometry, familiar to Riemann and his
followers!}

Dessins d'enfants (also known as bicolored maps) are bipartite graphs drawn on a smooth surface but they also possess manifold aspects. 
They are at the same time group theoretical, topological and algebraic objects and, as revealed by the author, they allow to stabilize the finite geometries attached to quantum contexts. There seems to exist a remarkable confluence between the so-called \lq magic' configurations of quantum observables found to illustrate the no-go theorems \`a la Kochen-Specker and the symmetries obeyed by the algebraic extensions over the field of rational numbers, a subject briefly advocated by Grothendieck as

{\it In the form in which Belyi states it, his result essentially says that every algebraic curve defined over a number field can be obtained as a covering of the projective line ramified only over the points $0$, $1$ and $\infty$. The result seems to have remained more or less unobserved. Yet it appears to me to have considerable importance. To me, its essential message is that there is a profound identity between the combinatorics of finite maps on the one hand, and the geometry of algebraic curves defined over number fields on the other. This deep result, together with the algebraic interpretation of maps, opens the door into a new, unexplored world - within reach of all, who pass by without seeing it}.

In Sec. \ref{dessins}, a brief account of the \lq technology' of dessins d'enfants is provided. Sec. \ref{Bellth} addresses the relation between Bell's theorem and some dessins \lq living' in the extension field $\mathbb{Q}(\sqrt{2})$. Sec. \ref{twoqubit} and \ref{threequbit}  deal about dessins attached to the two- and three-qubit Kochen-Specker theorem about contextuality. Then, in Sec. \ref{polygons}, it is shown that generalized polygons $GQ(2,2)$ and $GH(2,2)$, and their corresponding driving dessins, encode the commutation relations of two- and three-qubit sustems, respectively. As our last example, in Sec. \ref{sevencontext}, a dessin related to the contextuality of the six-qudit system, described in \cite{Cabello6}, is displayed.

\section{The manifolds traits of a \lq dessin d'enfant'}
\label{dessins}

\begin{figure}
\centering 
\includegraphics[width=6cm]{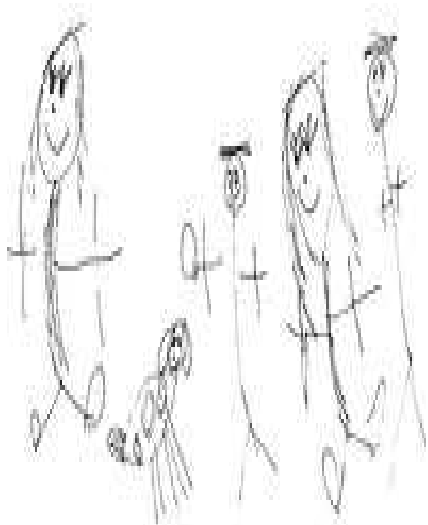}
\caption{A dessin d'enfant drawn at a kindergarten by a $5$-year old girl, http://www.parents.com/fun/arts-crafts/kid/decode-child-drawings/}
\label{fig1}
\end{figure}

I will explain that quantum contexts can be drawn as  Grothendieck's \lq dessins d'enfants'. A \lq true' dessin d'enfant is shown in Fig. 1. This section accounts for the mathematics of \lq false' \lq dessins d'enfants'. Of course, there are constraints that a child does not take care about: the \lq dessins' in question are connected, they are bipartite with black and white points and they are also chosen to be \lq clean', meaning that the valency of white vertices is $\le 2$. The last constraint can easily be removed --the valency of white vertices can be made arbitrary to correspond to an hypermap-- although this is not necessary for our quantum topic. Doing this, a \lq dessin' acquires a {\it topological genus} $g$ (which quantifies the number of holes on the smooth surface where it is drawn) such that $2-2g=B+W+F-n$, where $B$, $W$, $F$ and $n$ stands for the number of black vertices, the number of white vertices, the number of faces and the number of edges, respectively. 

Given a dessin $\mathcal{D}$  with $n$ edges labeled from 1 to $n$, one can recover the {\it combinatorial information} by associating with it a two-generator permutation group $P=\left\langle \alpha, \beta \right\rangle$ on the set of labels such that a cycle of $\alpha$ (resp. $\beta$) contains the labels of the edges incident to a black vertex (resp. white vertex) and by computing a passport \cite{Lando2004} in the form $[C_{\alpha},C_{\beta},C_{\gamma}]$, where the entry $C_i$, $i \in \{\alpha,\beta,\gamma\}$ has factors $l_i^{n_i}$, with $l_i$ denoting the length of the cycle and $n_i$ the number of cycles of length $l_i$.

Another observation made by Grothendieck is of utmost importance. The dessins are in one-to-one correspondence with conjugacy classes of subgroups of finite index of the triangle group $C_2^+=\left\langle  \rho_0,\rho_1,\rho_2|\rho_1^2=\rho_0\rho_1\rho_2=1 \right\rangle.$ The existence of dessins with prescribed properties can thus be straightforwardly checked from a systematic enumeration of conjugacy classses of $C_2^+$. Note that enumeration becomes tedius for large dessins since the number of dessins grows exponentially with the number of their edges. To proceed with the effective calculations of a dessin, one counts the cosets of a subgroup of $C_2^+$ and determine the corresponding permutation representation $P$ by means of the Todd-Coxeter algorithm implemented in an algebra software such as Magma.

Then, according to Belyi's theorem, a dessin may be seen as an {\it algebraic curve} over the rationals. Technically,
the {\it Belyi function} corresponding to a
dessin $\mathcal{D}$ is a rational function $f(x)$ of the complex variable $x$, of degree $n$,
such that (i) the black vertices are the roots of the equation $f(x)=0$ with the multiplicity of each root being equal to the degree of the corresponding (black) vertex, (ii) the white vertices are the roots of the equation $f(x)=1$ with the multiplicity of each root being equal to the degree of the corresponding (white) vertex,
(iii) the bicolored graph is the preimage of the segment $[0,1]$, that is $\mathcal{D}=f^{-1}([0,1])$,
(iv) there exists a single pole of $f(x)$, i.\,e. a root of the equation $f(x)=\infty$, at each face, the multiplicity of the pole being equal to the degree of the face, and, finally, 
(v) besides $0$, $1$ and $\infty$, there are no other critical values of $f$ \cite{Dessins2013,Lando2004}. This construction works well for small dessins  $\mathcal{D}$ but it becomes intractable for those with a high index $n$; however a complex algebraic curve is associated to every $\mathcal{D}$.

Last but not least, in many cases, one may establish a bijection between notable point/line incidence geometries $\mathcal{G}_
{\mathcal{D}}^i$ to a dessin $\mathcal{D}$, $i=1,\cdots,m$ with $m$ being the number of non-isomorphic subgroups $S$ of the permutation group $P$ of the dessin  that stabilize a pair of elements. We ask that every pair of points on a line shares the same stabilizer in $P$. Then, given 
a subgroup $S$ of $P$ which stabilizes a pair of points, we define the point-line relation on  $\mathcal{G}_{\mathcal{D}}$ such that two points will be adjacent if their stabilizer is isomorphic to $S$. A catalog of small finite geometries is given as tables 1 and 2 of \cite{Dessins2013}. Remarkably, most geometries derived so far have been found to rely on quantum contextuality, that is, the points of a $\mathcal{G}_{\mathcal{D}}$ correspond to quantum observables of multiple qubits and the lines are mutually commuting subsets of them.

\section{Bell's theorem with \lq dessins d'enfants'}
\label{Bellth}

John Bell: {\it First, and those of us who are inspired by Einstein would like this best, quantum mechanics may be wrong in sufficiently critical situations. Perhaps nature is not so queer as quantum mechanics. But the experimental situation is not very encouraging from this point of view\ldots
Secondly, it may be that it is not permissible to regard the experimental settings $a$ and $b$ in the analyzers as independent variables, as we did. We supposed them in particular to be independent of the supplementary variables $\lambda$, in that $a$ and $b$ could be changed without changing the probability distribution $\rho(\lambda)$\ldots Apparently seperate parts of the world would be deeply and conspirationnaly entangled, and our apparent free will would be entangled with them.
Thirdly, it may be that we have to admit causal influences do go faster than light\ldots
Fourthly and finally, it may be that Bohr's intuition was right --in that there is no reality below some \lq classical' \lq macroscopic' level. Then fundamental physical theory would remain fundamentally vague, until concepts like \lq macroscopic' could be made sharper than they are today.} \cite[p. 142]{Bell1981}

Bell's theorem is generally considered as as a proof of nonlocality, as in the third item of Bell's quote. But, as Bell's theorem is encompassed by Kochen-Specker theorem about contextuality, the second item of Bell's quote appears to be the most relevant, this alternative {\it rules out the introduction of exophysical automatons-with a random behavior-let alone observers endowed with free will. If you are willing to accept that option, then it is the entire
universe which is an indivisible, nonlocal entity.} \cite[p. 173]{Peres}.

Bell's theorem consists of an inequality that is obeyed by dichotomic classical variables but is violated by the (dichotomic) eigenvalues of a set of quantum operators. The simplest form of Bell's arguments \cite[p. 174]{Peres} makes use of four observables $\sigma_i$, $i = 1,2,3,4$, taking values in $\{-1,1\}$, of which Bob can measure $(\sigma_1,\sigma_3)$ and Alice $(\sigma_2,\sigma_4)$. One introduces the number
$$C=\sigma_1\sigma_2+\sigma_2\sigma_3+\sigma_3\sigma_4-\sigma_4\sigma_1=\pm 2$$
and observes the (so-called Bell/Clauser-Horne-Shimony-Holt (CHSH)/Cirel'son's) inequality \cite[p. 164]{Peres}
$$|\left\langle \sigma_1\sigma_2 \right\rangle+\left\langle \sigma_2\sigma_3 \right\rangle+\left\langle \sigma_3\sigma_4 \right\rangle-\left\langle \sigma_4\sigma_1 \right\rangle|\le 2,$$
where $\left\langle  \right\rangle$ here means that we are taking averages over many experiments. This inequality holds for any dichotomic random variables $\sigma_i$ that are governed by a joint probability distribution. Bell's theorem states that the aforementioned inequality is violated if one considers quantum observables with dichotomic eigenvalues. An illustrative example is the following set of two-qubit observables:
$\sigma_1=IX,~\sigma_2=XI,~\sigma_3=IZ,~\mbox{and}~\sigma_4=ZI,$
where $X$, $Y$ and $Z$ are the ordinary Pauli spin matrices and where, e.\,g., $IX$ is a short-hand for $I \otimes X$ (used also in the sequel).

The norm $||C||$ of $C$ (see \cite{Peres} or \cite{Dessins2013} for details) is found to obey $||C||=2 \sqrt{2}>2$, a maximal violation of the aforementioned inequality.

The point-line incidence geometry associated with our four observables is one of the simplest, that of a square -- Fig.\ref{fig1}a; each observable is represented by a point and two points are joined by a segment if the corresponding observables commute. It is  worth mentioning here that there are altogether 90 distinct squares among two-qubit observables and as many as
 $30240$ when three-qubit labeling is employed, each yielding a maximal violation of the Bell-CHSH inequality. 

\subsection*{Dessins d'enfants for the square  and their Belyi functions}

\begin{figure}
\centering 
\includegraphics[width=7cm]{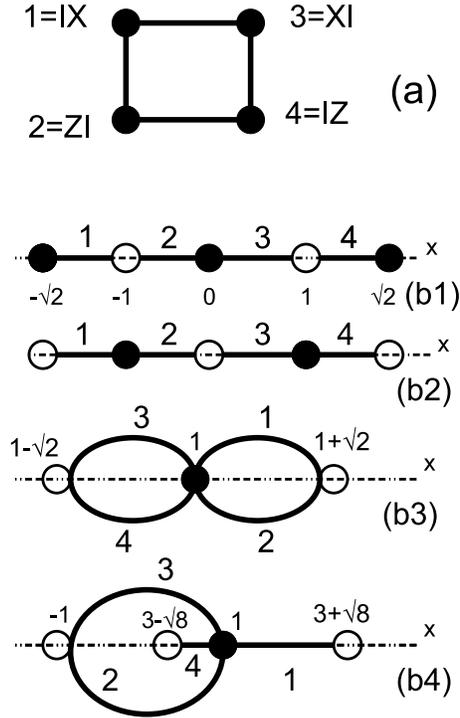}
\caption{A simple observable proof of Bell's theorem is embodied in the geometry of a (properly labeled) square (a) and four associated {\it dessins d'enfants}, ($b_1$) to ($b_4$). 
For each {\it dessin} an explicit labeling of its edges in terms of the four two-qubit observables is given. The (real-valued) coordinates of black and white vertices stem from the corresponding Belyi functions as explained in the main text.}
\label{fig2}
\end{figure}

The methodology described at Sec. \ref{dessins} was used to arrive at the result that the geometry of the square/quadrangle can be generated by four different {\it dessins}, $(b_1),\cdots, (b_4)$, associated with permutations groups $P$ isomorphic to the dihedral group $D_4$ of order $8$. Two of them, $(b_1)$ and $(b_2)$ are tree-like and the other two, $(b_3)$ and $(b_4)$, contain loops.

The first {\it dessin} ($b_1$) has the signature $s=(B,W,F,g)=(3,2,1,0)$ and the symmetry group $P=\left\langle (2,3),(1,2)(3,4)\right\rangle$ whose cycle structure reads $[2^1 1^2,2^2,4^1]$, i.e. one black vertex is of degree two, two black vertices have degree one, the two white vertices have degree two and the face has degree four. The corresponding Belyi function reads $f(x)=x^2(2-x^2)$, see \cite{Dessins2013} for details. The Belyi functions for the other cases $(b_2)$ to $(b_4)$ are $f(x)=(x^2-1)^2$, $f(x)=\frac{(x-1)^4}{4x(x-2)}$ and $f(x)=\frac{(x-1)^4}{16 x^2}$, respectively. Just observe that (critical) points of the dessin, where the derivative $f'(x)$ vanishes, correspond to black points where the valency is larger than one, that black point coordinates correspond to solutions of the equation $f(x)=0$, that white point coordinates correspond to the solution of the equation $f(x)=1$ and that the number of loops reflects in the number of poles of the corresponding Belyi function.

It is intriguing to see that all coordinates of a dessin live in the extension field $\mathbb{Q}(\sqrt{2})$ of the rational field $\mathbb{Q}$. Hence, a better understanding of the properties of the group of automorphisms of this field
 may lead to fresh insights into the nature of this important theorem of quantum physics. 

\section{Kochen-Specker theorem with \lq dessins d'enfants': two qubits}
\label{twoqubit}

{\it I am grateful to N.~D. Mermin for patiently explaining to me that ref. 11 [A. Peres, {\it Phys. Lett. A} {\bf 151}, 107 (1990); {\it Found. Phys.} {\bf 22}, 357 (1992)]) was a Kochen-Specker argument, not one about locality, as I had wrongly thought} \cite{PeresPOVM}.

Bell's theorem is a no-go theorem that forbids local hidden variable theories. Kochen-Specker theorem \cite{Koch1967} is stronger by placing new constraints in the permissible types of hidden variable theories. Kochen-Specker theorem forbids the simultaneous validity of the two statements (i), that all hidden variables have definite values at a given time (value definiteness) and (ii), that those variables are independent of the setting used to measure them (non-contextuality). Thus, quantum observables cannot represent the \lq elements of reality' of EPR paper \cite{Einstein1935}.

Kochen-Specker theorem establishes that even for compatible/commuting observables $A$ and $B$ with values $v(A)$ and $v(B)$, the equations $v(aA+bB)=av(A)+bv(B)$ ($a,b \in \mathbb{R}$) or $v(A)v(B)$ may be violated. The authors restricted the observables to a special class, viz. so-called yes-no observables, having only values $0$ and $1$, corresponding to projection operators on the eigenvectors of certain orthogonal bases of a Hilbert space.

\begin{figure}[ht]
\centering 
\includegraphics[width=6cm]{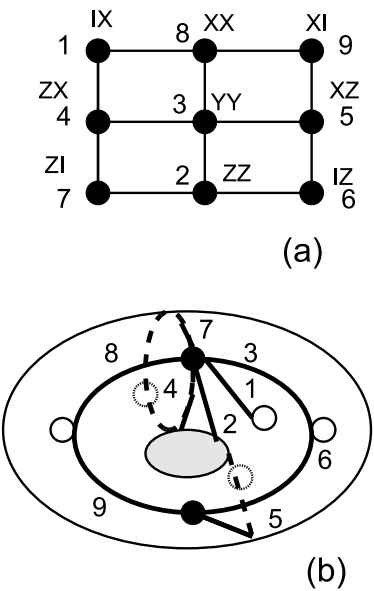}
\caption{A $3\times 3$ grid with points labeled by two-qubit observables ({\it aka} a Mermin magic square) (a) and a stabilizing {\it dessin} drawn on a torus (b).}
\label{fig3}
\end{figure}

One of the simplest types of violation is a set of nine two-qubit operators arranged in a $3 \times 3$-grid \cite{Mermin1993}. This grid is a remarkable one:  all triples of observables located in a row or a column are mutually commuting and have their product equal to   $+II$ {\it except for} the middle column, where $XX.YY.ZZ=-II$. Mermin was the first to observe that this is a Kochen-Specker (parity) type contradiction since the product of all triples  yields the matrix $-II$, while the product of corresponding eigenvalues is $+1$ (since each of the latter occurs twice, once in a row and once in a column) \cite[(b)]{Mermin1993}. Note that the Mermin square comprises a set of nine elementary squares/quadrangles that themselves constitute a proof of Bell's theorem, as shown at the preceeding section.

The Mermin `magic' square may be used to provide many contextuality proofs from the vectors shared by the maximal bases corresponding to a row/column of the diagram. The simplest, a so-called $(18,9)$ proof, ($18$ vectors and $9$ bases) has, remarkably, the orthogonality diagram which is itself a Mermin square ($9$ vertices for the bases and $18$ edges for the vectors) \cite[(c), eq. (6)]{Planat2013}.

\subsection*{A \lq dessin d'enfant' for the Mermin square}

Mermin square is shown Fig. \ref{fig3}a. One can recover this geometry with a genus one {\it dessin}, with signature $(2,5,2,1)$, as shown in Fig. \ref{fig3}b. The corresponding permutation group is 
$P=\left\langle (1,2,4,8,7,3)(5,9,6),(2,5)(3,6)(4,7)(8,9)\right\rangle \cong \mathbb{Z}_3^2 \rtimes \mathbb{Z}_2^2$, having the cycle structure  $[6^13^1,2^41^1,6^13^1]$.
This {\it dessin} lies on a Riemann surface that is a torus (not a sphere $\hat{\mathbb{C}}$), being thus represented by an elliptic curve. The topic is far more advanced and we shall not pursue it in this paper (see, e.\,g., \cite{Girondo} for details).
The stabilizer of a pair of edges of the {\it dessin} is either the group $\mathbb{Z}_2$, yielding Mermin's square $M_1$ shown in Fig \ref{fig3}a, or the group $\mathbb{Z}_1$, giving rise to a different square $M_2$ from the maximum sets of mutually non-collinear pairs of points of $M_1$. The union of $M_1$ and $M_2$ is the Hesse configuration.

\section{Kochen-Specker theorem with \lq dessins d'enfants': three qubits}
\label{threequbit}

Poincar\'e wrote: {\it Perceptual space is only an image of geometric space, an image altered by a sort of perspective \cite[p. 342]{Flanagan}}
and Weyl wrote: {\it In this sense the projective plane and the color continuum are isomorphic with one another \cite[p. 343]{Flanagan}.}

\begin{figure}[ht]
\centering 
\includegraphics[width=6cm]{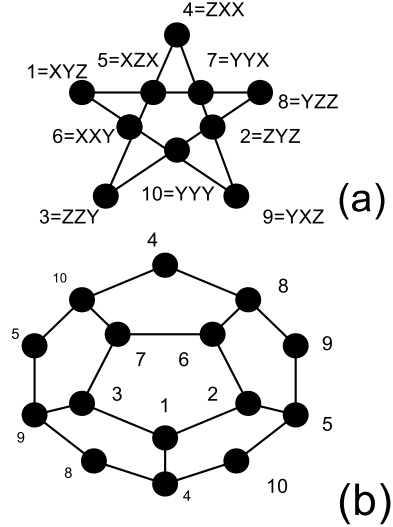}
\caption{(a) A Mermin pentagram $\bar{\mathcal{P}}$ and (b) the embedding of the associated Petersen graph $\mathcal{P}$ on the real projective plane as a hemi-dodecahedron.}
\label{fig4}
\end{figure}

Color experience through our eyes to our mind relies on the real projective plane $\mathbb{R}\mathbb{P}^2$ \cite{Flanagan}. Three-qubit contextuality also relies on $\mathbb{R}\mathbb{P}^2$ thanks to a Mermin `magic' pentagram, that for reasons explained below in (i) we denote $\bar{\mathcal{P}}$ (by abuse of language because we are at first more interested to see the pentagram as a geometrical configuration than as a graph). 
One such a pentagram is displayed in Fig. 4a. It consists of a set of five lines, each hosting four mutually commuting operators and any two sharing a single operator. The product of operators on each of the lines is $-III$, where $I$ is the $2 \times 2$ identity matrix. It is impossible to assign the dichotomic truth values $\pm 1$ to eigenvalues while keeping the multiplicative properties of operators so that the Mermin pentagram is, like its two-qubit sibling, `magic', and so contextual \cite[a]{Mermin1993},\cite[(b) and c)]{Planat2013}.

\begin{figure}[ht]
\centering 
\includegraphics[width=6cm]{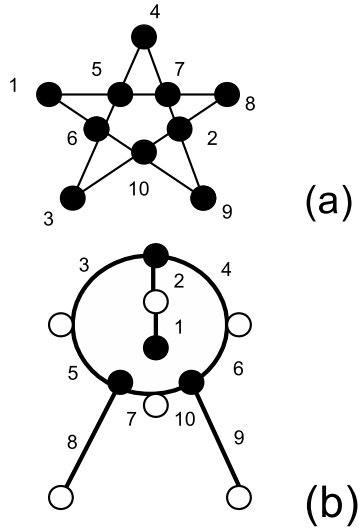}
\caption{(a)  Mermin pentagram $\bar{\mathcal{P}}$ and (b) a generating dessin.}
\label{fig5}
\end{figure}

Let us enumerate a few remarkable facts about a pentagram. 

(i) The graph $\bar{\mathcal{P}}$ of a pentagram is the complement of that of the celebrated Petersen graph $\mathcal{P}$. One noticeable property of $\mathcal{P}$ is to be the smallest bridgeless cubic graph with no three-edge-coloring. The Petersen graph is thus not planar, but it can be embedded without crossings on $\mathbb{R}\mathbb{P}^2$ (one of the simplest non-orientable surfaces), as illustrated in Fig 4b. 

(ii) The Petersen graph may also be seens as the complement of the Desargues configuration $10_3$ (a celebrated projective geometry that has ten lines with three points and ten points each of them incident with three lines), see \cite[Fig. 11]{Dessins2013}. 

(iii) There exist altogether $12096$ three-qubit Mermin pentagrams, this number being identical to that of automorphisms of the smallest split Cayley hexagon $GH(2,2)$ -- a remarkable configuration of $63$ points and $63$ lines \cite[(b)]{Planat2013} pictued in Fig. 7a.

(iv) Now comes an item close to the {{\it it from bit} perspective. The Shannon capacity of a graph is the maximum number of $k$-letter messages than can be sent through a channel without a risk of confusion.  The Shannon capacity of $\mathcal{P}$ is found to be optimal and equal to $4$, much larger than that $\sqrt{5}$ of an ordinary pentagon.


(v) Finally, the pentagram configuration in Fig. 5a may be generated/stabilized by a 'dessin d'enfant' on the Riemann sphere, having permutation group isomorphic to the alternating group $A_5$ and cycle structure $[3^2 1^1,2^41^2,5^2]$,  as shown in Fig. 5b. The stabilizer of a pair of edges of the {\it dessin} is either the group $\mathbb{Z}_1$ giving rise to the Mermin's pentagram, or the group $\mathbb{Z}_2$ giving rise to the Petersen graph.

\section{Dessins d'enfants and generalized polygons}
\label{polygons}

Jacques Tits: {\it  I would say that mathematics coming from physics is of high quality. Some of the best results we have in mathematics have been discovered by physicists. I am less sure about sociology and human science \cite{Tits}.}

\begin{figure}[ht]
\centering 
\includegraphics[width=8cm]{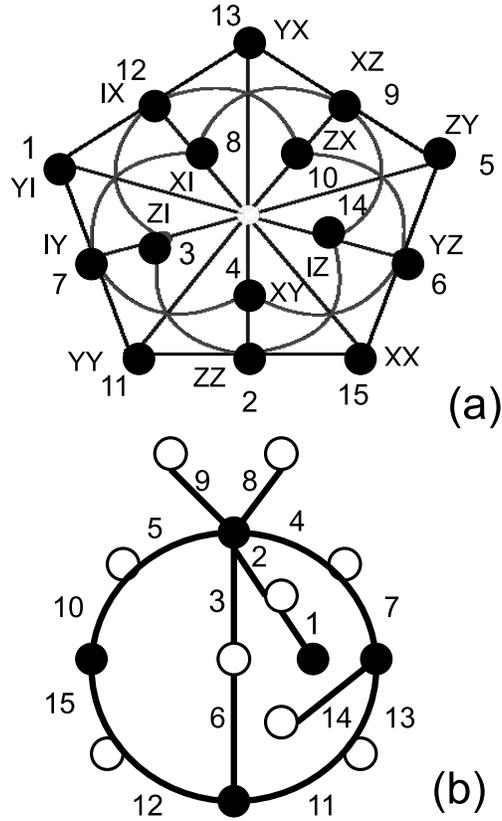}
\caption{The generalized quadrangle $GQ(2,2)$ (a) with its points labeled by the elements of the two-qubit Pauli group and a stabilizing {\it dessin} (b).}
\label{fig11}
\end{figure}

Jacques Tits discovered generalized polygons (also called generalized 
$n$-gons). A generalized polygon is an incidence structure between a discrete set of points and lines whose incidence graph has diameter $n$ (the maximum eccentricity of any vertex) and girth $2n$ (the length of a shortest cycle). A generalized polygon of order $(s,t)$ has every line containing $s+1$ points and ever point lying on $(t+1)$ lines. Remarkably, the generalized $4$-gon/quadrangle of order $(2,2)$, namely $GQ(2,2)$ controls the commutation structure of the $15$ two-qubit observables \cite[(a)]{Planat2013} and the generalized $6$-gon/hexagon $GH(2,2)$ does the job for the $63$ three-qubit observables \cite[(b)]{Planat2013}.

An important concept pertaining to generalized polygons is that of a geometric hyperplane. A geometric hyperplane of a generalized polygon is a proper subspace meeting each line at a unique point or containing the whole line. The substructure of a polygon of order $(2,t)$ highly relies on its hyperplanes in the sense that one \lq adds' any two of them to form another geometric hyperplane. The \lq addition' law in question is nothing but \lq the complement of the symmetric difference' of the two sets of points involved in the pair of selected hyperplanes. There exists three kinds of hyperplanes in $GQ(2,2)$ one of them being the Mermin square described in Sec. \ref{twoqubit}. The structure of hyperplanes of $GH(2,2)$ is of utmost importance to describe the three-qubit Kochen-Specker theorem as described exhaustively in \cite[(b)]{Planat2013}.

Next, we find that generalized polygons are induced/stabilized by  dessins d'enfants. As for the case of the geometry of the square, a selected geometry $\mathcal{G}$ may be induced/stabilized by many dessins $\mathcal{D}_i$, i.e. the correspondance $f:\mathcal{D}_i \rightarrow \mathcal{G}$ is non injective. Moreover, the number of dessins grows exponentially with the number of their edges so that the systematic search of all maps $f$ may become tedious. To simplify the search of a solution $f$ inducing a selected $\mathcal{G}$ (such as a generalized polygon) one restricts the search to subgroups of the cartographic group $C_2^+$  (go back to Sec. 2 for the definition).

\subsection*{A dessin d'enfant for the generalized quadrangle $GQ(2,2)$}

The generalized quadrangle $GQ(2,2)$ encodes the commutation relations of two-qubit operators as shown in Fig. 6a \cite[(a)]{Planat2013}.

A dessin stabilizing the generalized quadrangle $GQ(2,2)$ may be obtained by studying the conjugacy classes of the subgroup $C_2^+/[\rho_2^4=1]$, whose permutation representation of the cosets is isomorphic to the symmetry group of $GQ(2,2)$, that is the symmetric group $S_6$. 
This is shown in Fig. 6. One finds that the dessin in Fig. 6b has two types of stabilizers for a pair od edges, one isomorphic to $\mathbb{Z}_2^5$ and inducing $GQ(2,2)$ and the other one isomorphic to $\mathbb{Z}_6$ and inducing the complement of $GQ(2,2)$.

\subsection*{A dessin d'enfant for the generalized hexagon $GH(2,2)$}

The generalized hexagon $GH(2,2)$ encodes the commutation relations of three-qubit operators as shown in Fig. 7a \cite[(b)]{Planat2013}.

\begin{figure}[ht]
\centering 
\includegraphics[width=9cm]{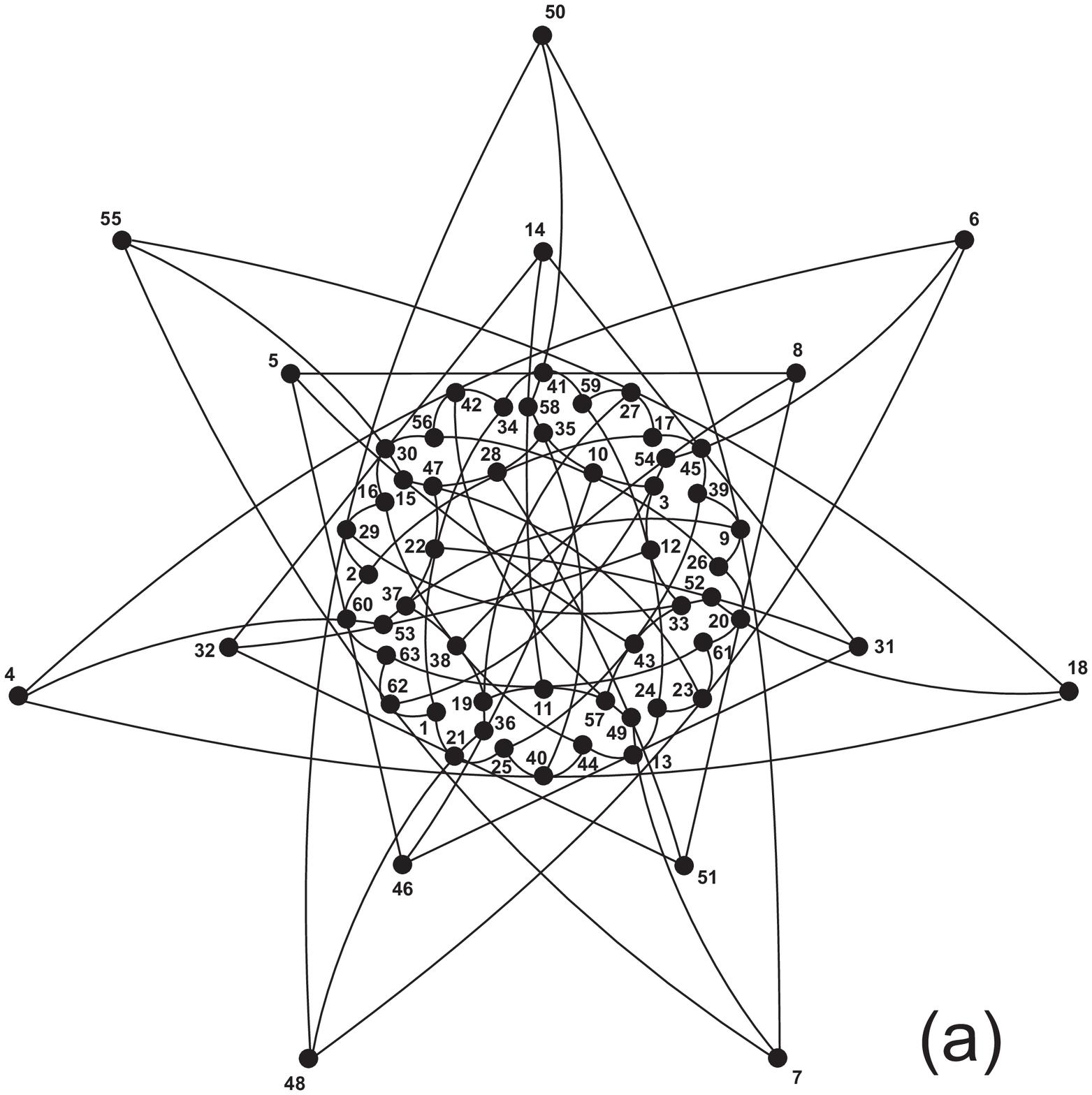}
\includegraphics[width=6cm]{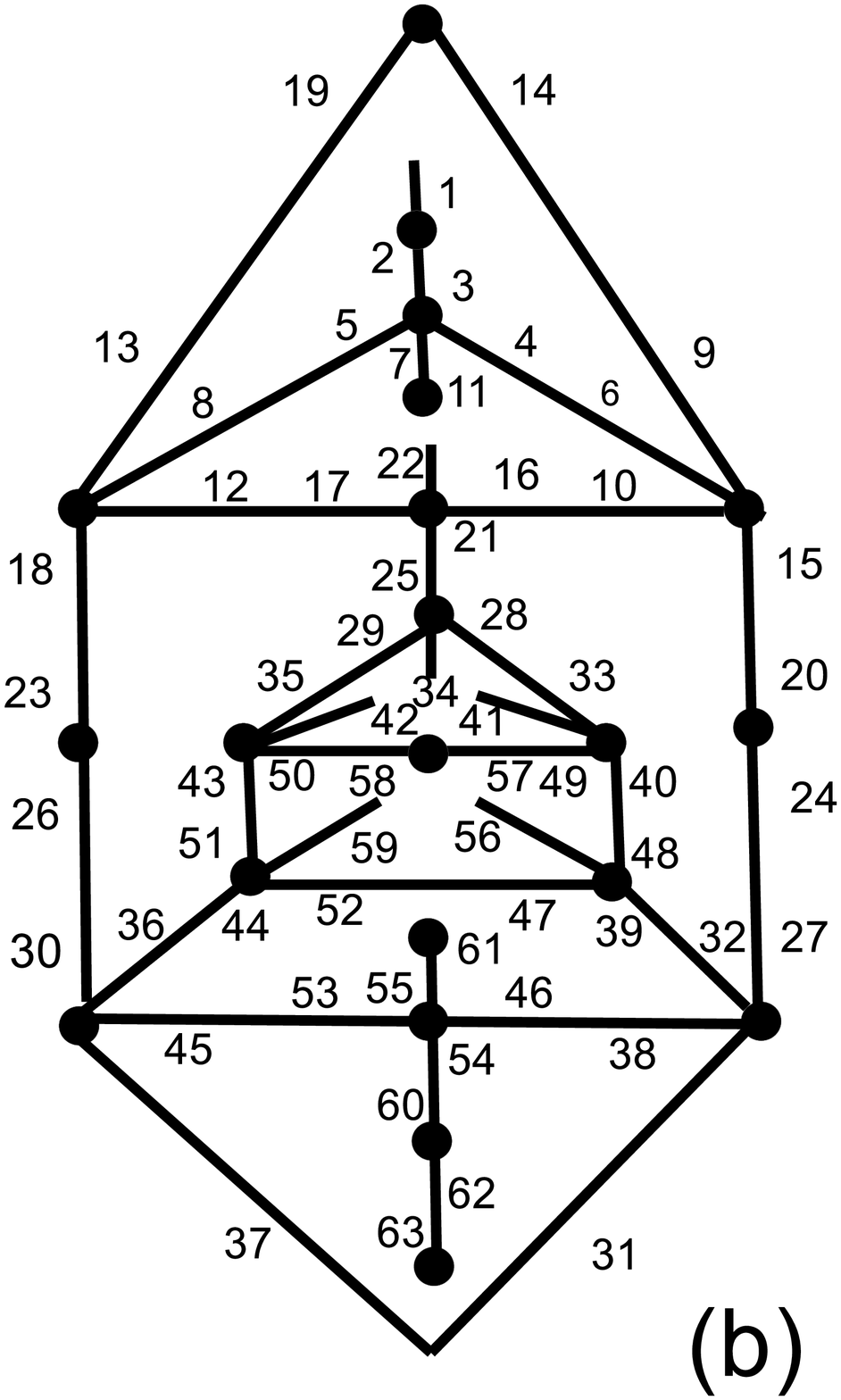}
\caption{(a) The hexagon $GH(2,2)$ and (b) the {\it dessin} for its collinearity graph. To simplify the drawing, white points are not shown but half-edges are labelled.}
\label{dessinHexagon}
\end{figure}

A dessin stabilizing the generalized hexagon $GH(2,2)$ may be obtained by studying the conjugacy classes of a  subgroup of $C_2^+$ whose finite representation is that of the symmetry group of the hexagon. Then one selects the dessins of permutation group $P$ isomorphic to the wreath product $S_3 \wr S_3$ (of order $1296$). Remarkably, one finds only two dessins satisfy these requirements, one is of genus $0$ and induces $GH(2,2)$ as shown in Fig. 7 and the other one (not shown) is of genus $1$ (drawn on a torus) and induces the dual of $GH(2,2)$.

\section{A dessin d'enfant for six-qudit contextuality }
\label{sevencontext}

\begin{figure}[ht]
\includegraphics[width=8cm]{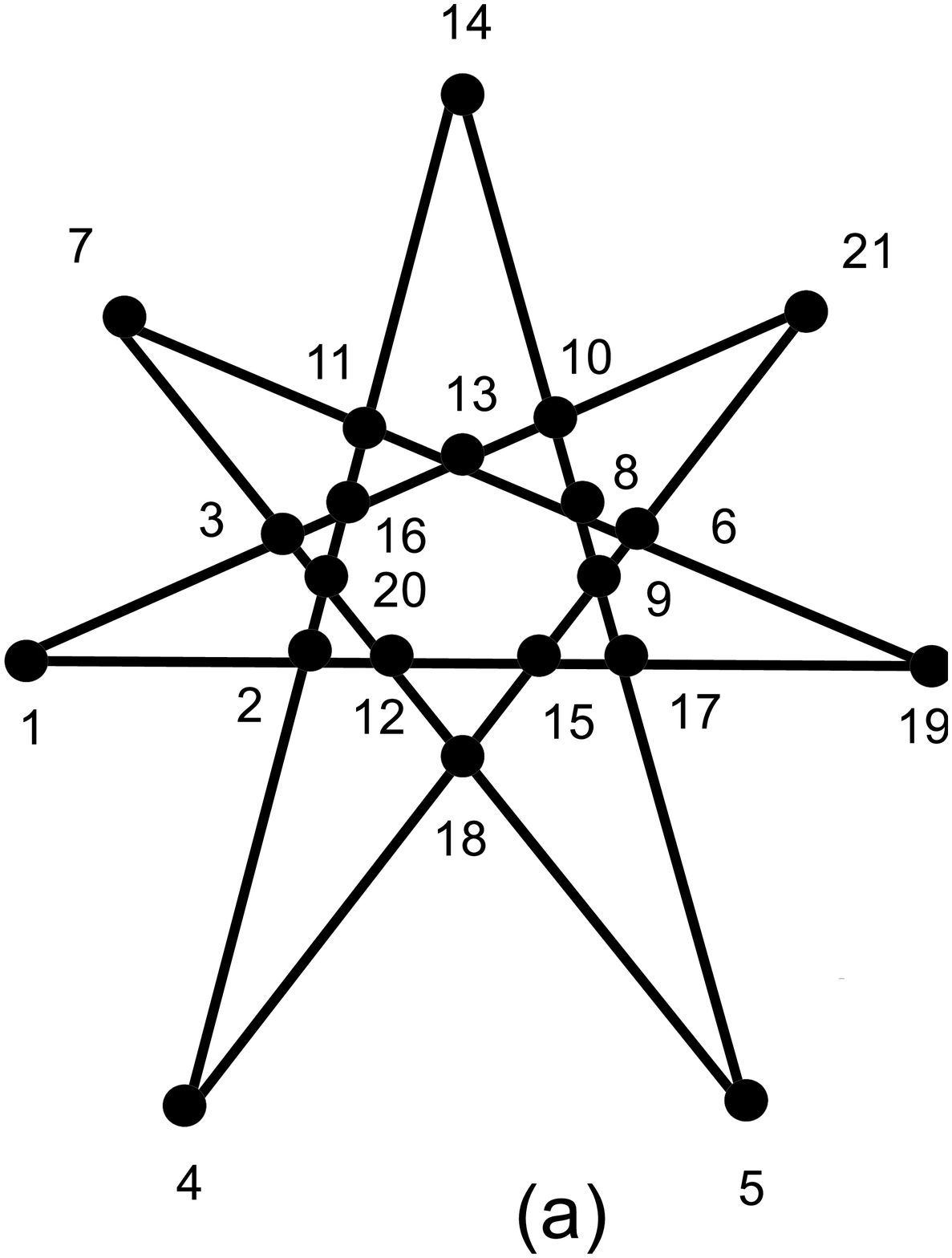}
\includegraphics[width=8cm]{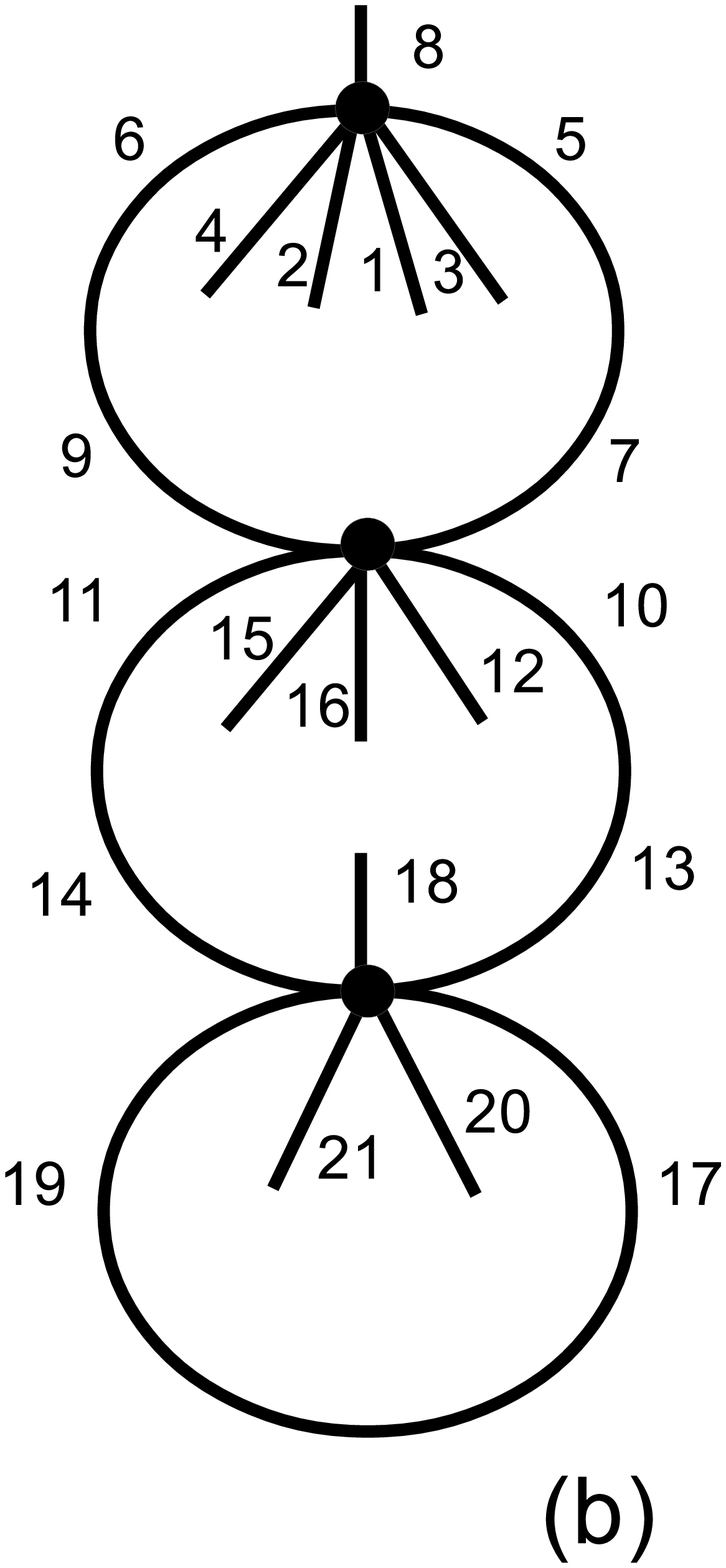}
\caption{ (a) The seven-context geometry of six-qudit contextuality and (b) a dessin stabilizing the $35$-triangle geometry lying in (a). To simplify the drawing, white points are not shown but half-edges are labelled.}
\label{dessinheptagram}
\end{figure}

Recently, a remarkable (minimal) Kochen-Specker configuration built from seven contexts and $21$ rays, belonging to a six-qudit system, has been built \cite{Cabello6}. The set in question is an heptagram (see Fig. 8a) in which the seven lines are hexads of mutually orthogonal vectors (they are not quantum observables as was the case at the previous sections). It is straightforward to check (with a computer) that the collinearity graph of this geometry contains two kinds of maximal cliques, the seven hexads just mentioned and, in addition, $35$ triangles. The graph of the latter $35$-triangle geometry is nothing but the line graph of the complete graph $K_7$ [the line graph of the complete graph $K_5$ is the Desargues configuration that occured at Sec. \ref{threequbit}, item (ii)]. The symmetry group of the $35$-triangle geometry, as that of the heptagram, is the seven-letter symmetric group $S_7$. Starting from a finite representation of $S_7$ that underlies a subgroup of the cartographic group $C_2^+$, it is not difficult to find a dessin stabilizing the aforementioned $35$-triangle geometry, as shown in Fig. 8b. Once again, it has been shown that quantum contexts are intimately related to dessins d'enfants.

\section{Conclusion}

It was not anticipated by his creator that geometries induced by \lq dessins d'enfants' would so nicely fit the drawings/geometries underlying quantum contextuality. I believe that this key observation opens new vistas for the interpretation of quantum measurements in terms of algebraic curves over the rationals. The symmetries of general dessins rely on a fascinating group called the universal (or absolute) Galois group over the rationals, a still rather mysterious object. May be the present work gives some substance to an hidden variable interpretation of the quantum world, as hoped by Einstein, but in a subtle way. We just mentioned in passing (in Sec. \ref{polygons}) that the geometries relevant to quantum contexts have a rich substructure of hyperplanes that also has to be incorporated in the new design. More details will be given at a next stage of our research.




\end{document}